\documentclass{article}

\def \br{{\bf r}}

\def \la{\label}
\def \dd{\hbox{d}}
\newcommand{\be}{\begin{equation}}
\newcommand{\ee}{\end{equation}}
\newcommand{\beq}{\begin{eqnarray}}
\newcommand{\eeq}{\end{eqnarray}}

\usepackage{amsmath}
\usepackage{graphicx}
\usepackage{cite}
\begin{document}

\author{Marek Napi\'orkowski \footnote{Marek.Napiorkowski@fuw.edu.pl} and Jaros{\l}aw Piasecki \\
Institute of Theoretical Physics, Faculty of Physics \\ University of Warsaw\\
 Ho\.za 69, 00-681 Warsaw, Poland}
\title{The bulk correlation length and the range of thermodynamic Casimir forces at Bose-Einstein condensation}
\maketitle
\begin{abstract}
The relation between the bulk correlation length and the decay length of thermodynamic Casimir forces is investigated microscopically in two three-dimensional systems undergoing Bose-Einstein condensation: the perfect Bose gas and the imperfect mean-field Bose gas. For each of these systems, both lengths diverge upon approaching the corresponding condensation point from the one-phase side, and are proportional to each other. We determine the proportionality factors and discuss their dependence on the boundary conditions. The values of the corresponding critical exponents for the decay length and the correlation length are the same, equal to $1/2$ for the perfect gas, and $1$ for the imperfect gas.
\end{abstract}
\newpage

\section{Introduction}
The thermal Casimir effect has been the subject of intensive theoretical \cite{K1994,BTD2000,Bar1983,FdG1978,Ha2011,G2009,VGMD2009,MB2006,MZ2006,GD2006,KG1999,BU1998,FU1992,SY1981,KD1992,S1993,D1998,DR2001,RS2002,DDG2006,SD2008,GD2008,DS2011,JOP2009} and experimental research \cite{HHGDB2009,HHG2008,RBM2007,FYP2005} in condensed matter physics. One typically considers a strongly fluctuating system contained  between two parallel plane walls. The approach of the system's thermodynamic state  towards its bulk critical point is accompanied by the appearance of increasingly long-range so-called thermodynamic Casimir forces acting between the walls.  At the same time the vicinity of the bulk critical state induces order-parameter correlations of macroscopic range. It is thus natural to expect a close relation between the Casimir effect  and the nature of these correlations in the vicinity of the critical state. In particular, the theory of finite-size scaling in the critical region \cite{K1994,BTD2000,Bar1983} points at proportionality between the bulk correlation length and the decay length of the Casimir forces. However, it does not  specify the value of their ratio. Our purpose is to determine this ratio within the framework of microscopic analysis. \\

  We concentrate our attention on two three-dimensional systems, each undergoing Bose-Einstein condensation: the perfect Bose gas and the imperfect mean-field  Bose gas. Our choice is dictated by the fact that a straightforward rigorous and microscopic analysis can be performed for each system. We examine the relationship between the properties of the bulk correlation length, and the decay length characterizing the thermodynamic Casimir forces at the approach to the condensation point. The decay length depends on the type of the boundary conditions imposed at the walls while the bulk correlation length is independent of the boundary conditions. Near the condensation point the behavior of the  thermodynamic Casimir force is described by the scaling function from which one can read off the unique expression  for the decay length. Its relation to the bulk correlation length can be established only after an independent calculation of the latter quantity is performed.  One can th!
 us rephrase our problem as determining  the bulk correlation length and the Casimir force decay length in two independent calculations for two systems and establishing the exact relation between these two quantities for each system. \\

The study of the correlation length requires the knowledge of the number density 
$n_{2}(|\br_{1}-\br_{2}|)$ of pairs of bosons situated at distance $ |\br_{1}-\br_{2}|$. The density 
$n_{2}(|\br_{1}-\br_{2}|)$ tends to $\rho^2$ for $|\br_{1}-\br_{2}|\to\infty$, where $\rho$ is the one-particle density. 
The large distance behavior of the correlation function 
\be
\label{cprrf}
\chi (r) = n_{2}(r) - \rho^2
\ee
changes qualitatively when the system undergoes Bose-Einstein condensation. In the absence of condensate, the behavior of the correlation function  $\chi(r)$ is dominated by the exponential decay $\sim \exp(-r/\xi)$, whereas in the two-phase region one observes a slow non-integrable power law decay. This qualitative change is reflected by the fact that in the one-phase region the correlation length $\xi$ governing the exponential decay tends to infinity upon approaching the condensation point. 

A quite similar behavior is observed for the thermodynamic Casimir force coupling two parallel plane walls separated by distance $D$.
One way to evaluate the force is to determine, for a given temperature $T$ and chemical potential $\mu$, the excess free energy density per unit wall area
\be
\label{gcdensity}
\omega_{s}(T ,D,\mu) =    \omega (T,D,\mu) - D\,\omega_{b}(T,\mu)  \quad. 
\ee
Here $\omega (T,D,\mu)$ is the total grand canonical free-energy density per unit wall area, and $\omega_{b}(T,\mu)$ denotes the bulk grand canonical free-energy density evaluated in the thermodynamic limit. 
The calculation of  $\omega_{s}(T,D,\mu)$ requires the thorough analysis of finite size effects under specific boundary conditions. 

By definition, the Casimir force is given by the formula
\be
\label{FD}
F(T,D,\mu) = -\frac{\partial \omega_{s}(T,D,\mu)}{\partial D} \quad.
\ee
Much like the correlation function, the $D$-dependence of the thermodynamic Casimir force $F(T,D,\mu)$ in the one-phase region is dominated by the exponential decay $\sim \exp(-D/\kappa)$ for $D\to\infty$. 
However, its range $\kappa$ approaches infinity in the vicinity of the condensation point announcing the appearance of a new phase. 
In the presence of condensate the force $F(T,D,\mu)$  decays according to a power law $ D^{-3}$.

As already mentioned, the object of the present study is to compare the expressions for the correlation length $\xi$ and the 
range $\kappa$ of the thermodynamic Casimir  force for perfect and imperfect Bose gas when the corresponding condensation points are approached  from the one-phase regions. 

\section{Perfect Bose gas}
In the case of a perfect Bose gas  the correlation function $\chi_{0}(r)$ in the one-phase region (no condensate) 
is given by (see the excellent review \cite{ZUK77})
\be
\label{ntwo}
\chi_{0}(r) = [ F(r,\alpha , T)]^2 \quad, 
\ee
where 
\be
\label{defF}
\lambda^3\, F(r,\alpha , T ) = \sum_{j=1}^{\infty}\frac{1}{j^{3/2}}\exp\left[ -\alpha\, j -\frac{\pi r^2 }{j\lambda^2} \right] \quad. 
\ee
Here $\lambda = h/\sqrt{2\pi mk_{B}T}$ is the thermal de Broglie wavelength  and $\alpha = (-\mu )/k_{B}T$. 
 
A particularly useful representation of series (\ref{defF}) has the form \cite{ZUK77}
\be
\label{defFF}
\lambda^3\, F(r,\alpha , T) = \frac{\lambda}{r}\exp\left( -2\frac{\sqrt{\pi\alpha}\, r}{\lambda}  \right) + 
\sum_{s=1}^{\infty}\frac{\lambda}{r}\exp\left[ -A^{+}(s)\frac{r}{\lambda} \right]
2 \cos\left[ -A^{-}(s)\frac{r}{\lambda} \right] 
\ee
with
\be
 A^{\pm}(s) = \sqrt{2\pi}(\alpha^2+4\pi^2 s^2)^{1/4}\left[ 1 \pm \frac{\alpha}{(\alpha^2+4\pi^2 s^2)^{1/2}} \right]  \quad. \nonumber
\ee

It follows from (\ref{ntwo}) and (\ref{defFF}) that when the critical value $\mu_{c,0}=0$ of the chemical potential is approached from below, 
and thus $\alpha\to 0$, the correlation function $\chi_{0}(r)$ at large distances decays according to the asymptotic formula 
\be
\label{decay0}
\lambda^{6}\chi_{0} (r) |_{r\gg \lambda} \cong  \left(\frac{\lambda}{r}\right)^2\exp\left( -4\frac{r}{\lambda}\sqrt{\pi\alpha }  \right) =
\left( \frac{\lambda}{r}\right)^2 \exp\left( - \frac{r}{\xi_{0}} \right) \quad. 
\ee
From (\ref{decay0}) we find that the bulk correlation length $\xi_{0}$ of a perfect Bose gas is given by
\be
\label{corrrange0}
\xi_{0}(\mu ) =  \frac{\lambda}{4\sqrt{\pi\alpha}} = \frac{\lambda}{4}\sqrt{\frac{k_{B}T}{\pi (-\mu )}} = \frac{h}{4\pi\sqrt{2m}}\frac{1}{\sqrt{-\mu}} \quad.
\ee
When $\mu <0$ approaches its critical value $\mu_{0,c}=0$,  the correlation length $\xi_{0}(\mu , T)$ diverges  as 
${\cal A}_{0} (-\mu )^{-\nu_{0}}$ with the critical exponent $\nu_{0} = 1/2$, and the temperature independent amplitude 
\be
\label{amplitude0}
{\cal A}_{0}= \frac{h}{4\pi\sqrt{2m}} \quad.
\ee

It is our purpose here to compare the divergence of the correlation length $\xi_{0}(\mu)$  with that of the 
decay length characterizing the thermodynamic Casimir force (\ref{FD}). 

The Casimir force $F(T,D,\mu)$ for a perfect Bose gas has been evaluated by various methods; inter alia  by Symanzik \cite{SY1981} via field-theoretic approach, by Krech and Dietrich 
\cite{KD1992}, Gr\"uneberg and Diehl \cite{GD2008}, and  by Zagrebnov and Martin \cite{MZ2006} by statistical physics methods for various boundary conditions: periodic (per), Dirichlet (D), 
and Neumann (N). The Robin boundary conditions were discussed by Romeo and Saharian \cite{RS2002}, and by Diehl and Schmidt \cite{SD2008,DS2011}. The Casimir force in the spherical model 
with periodic boundary conditions was discussed by Sachdev \cite{S1993}, and Danchev \cite{D1998}. In the one-phase region 
($\mu<0$) the large $D$-dependence of the thermodynamic Casimir force is dominated by the exponential decay $\sim \exp(-D/\kappa_{0})$,  where the decay length $\kappa_{0}$ for periodic  boundary 
conditions is given by (see equation (21) in Ref.\cite{MZ2006})
\be
\label{rangeF0}
\kappa_{0,per} (\mu) = \frac{\lambda}{4}\,\sqrt{\frac{k_{B}T}{\pi (-\mu )}} \quad. 
\ee
The decay length $\kappa_{0,per} (\mu)$ coincides thus exactly with the correlation length $ \xi_{0}(\mu )$. 

 In the case of Dirichlet and Neumann boundary conditions one obtains $\kappa_{0,D} = \kappa_{0,N} = \kappa_{0,per}/2$. The equalities 
$ \xi_{0}(\mu ) = \kappa_{0,per} (\mu)= 2\kappa_{0,D} = 2\kappa_{0,N}$  show clearly that the same mechanism is responsible for the appearance of the thermodynamic Casimir force between the walls and for building up macroscopic range of correlations  near the condensation point. 

\section{Imperfect Bose gas}

The Hamiltonian of the imperfect Bose gas has the form
\be
\label{impham}
H_{imp} = H_{0} + \frac{a N^2}{2V} \quad,
\ee
where $H_{0}$ is the kinetic energy operator. The term $[ a N^2/2V ] $, where $a=\int \dd \mathbf{r} \; \Phi(r)$ is a positive constant, takes into account the 
potential energy $\Phi(r)$ of repulsive interparticle interactions in the mean-field approximation (see Appendix, Eq.(\ref{mfham}) ). In fact, $a/V>0$ can be looked upon as the constant mean-field potential energy per pair of bosons. Since the mean-field theory can be obtained from the Kac's scaling of long-range interparticle potential \cite{L1986} it is worthwhile to note that the Casimir forces in the presence of long-range interactions in the spherical model were discussed by Danchev and Rudnick \cite{DR2001}, and by Danchev, Diehl, and Gr\"uneberg \cite{DDG2006}.  

Our original derivation of the two-particle density $n_{2,imp}(|\br_{1}-\br_{2}|)$ for an imperfect Bose gas
is presented in the Appendix. We show therein that this problem can be reduced to that of a perfect Bose gas with Hamiltonian $H_{0}$, 
and that equations (\ref{ntwo}) and (\ref{defF}) continue to hold provided the parameter $\alpha = - \mu/(k_{B}T)$ is replaced by 
\be
\label{alfamf}
\bar{\alpha}(T,\mu) = -\frac{\nu(T,\mu)}{k_{B}T} =  - \left[\frac{\mu - a\rho(T,\mu)}{k_{B}T}\right] \quad, 
\ee
The mean-field critical value of the chemical potential is positive and given by
\be
\label{mucmf}
\mu_{imp,c} = a\rho_{0,c} \quad,
\ee
where $\rho_{0,c}$ denotes the critical density of  a perfect Bose gas;  $\lambda^{3} \rho_{0,c}= \zeta(3/2)=2.612$ .  This fact is a direct consequence
of the implicit equation 
\be
\label{implicit}
\rho (T,\mu) = \rho_{0}(T,\nu(T,\mu)) = \rho_{0}[T, \mu - a\rho(T,\mu)]
\ee
which  characterizes the equilibrium state of an imperfect Bose gas ( see e.g. \cite{BLS1984}).

Consequently,  the decay of the correlation function $\chi_{imp}(r)$ in the region $\mu < \mu_{imp,c}$ is governed by the exponential law
\begin{eqnarray}
\label{decaymf}
\lambda^6 \, \chi_{imp} (r) |_{r\gg\lambda} \cong \left(\frac{\lambda}{r}\right)^2\exp\left( -4\frac{ r}{\lambda} 
\sqrt{\pi \bar{\alpha}}\ \right) \nonumber = \\
 \left(\frac{\lambda}{r}\right)^2\exp\left( -4\frac{ r}{\lambda} \sqrt{-\frac{\pi [\mu - a\rho(T, \mu)]}{k_{B}T}}\, \right) 
= \left(\frac{\lambda}{r}\right)^2\exp(-r/\xi_{imp})
\end{eqnarray}
(compare with (\ref{decay0})).

In order to investigate the divergence  of the correlation length  $\xi_{imp}$ upon approaching the condensation point we have to analyze the behavior of the function $\nu(T,\mu) = \mu - a\rho(T,\mu)$  for $\mu \to \mu_{imp,c}$. 
The perfect gas density for $\nu<0$ is given by the series
\be
\label{rhozero}
\lambda^3 \rho_{0}(T,\nu) = \sum_{q=1}^{\infty}\frac{1}{q^{3/2}}\exp ( \nu q /k_{B}T) \quad. 
\ee
Equations (\ref{implicit}) and (\ref{rhozero}) imply the following formulae
\be
\label{derivative1}
 \frac{\partial\rho}{\partial\mu} = \left(1+a\frac{\partial\rho_{0}}{\partial\nu}\right)^{-1}\frac{\partial\rho_{0}}{\partial\nu}
= \frac{1}{a}\left[ 1 -  \left(1+a\frac{\partial\rho_{0}}{\partial\nu}\right)^{-1} \right] \quad,
\ee
\be
\label{derivative2}
 \frac{\partial^{2}\rho}{\partial\mu ^{2}} =  \left(1+a\frac{\partial\rho_{0}}{\partial\nu}\right)^{-3} \frac{\partial^{2}\rho_{0}}{\partial\nu ^{2}} \quad.
\ee
Using the standard notation for the Bose functions 
\be
\label{Bosef}
g_{n}(\bar{\alpha}) = \sum_{q=1}^{\infty}\frac{\exp (- \bar{\alpha}q)}{q^{n}}
\ee
we find the following relations
\[ \lambda^3 \rho_{0}(\nu) = g_{3/2}(\bar{\alpha})  \quad, \] 
\[ k_{B}T \lambda^3\frac{\partial\rho_{0}}{\partial\nu} =  g_{1/2}(\bar{\alpha}) \quad, \] 
\[( k_{B}T)^2 \lambda^3\frac{\partial^{2}\rho_{0}}{\partial\nu ^{2}} = g_{-1/2}(\bar{\alpha}) \quad. \] 
The asymptotic behavior of $g_{1/2}(\bar{\alpha})$, and  $g_{-1/2}(\bar{\alpha})$ for $\bar{\alpha}\to 0$ reads \cite{ZUK77}
\be
\label{divergence}
 g_{1/2}(\bar{\alpha}) \cong \sqrt{\frac{\pi}{\bar{\alpha}}} \; , \;\;\;\; g_{-1/2}(\bar{\alpha}) \cong \frac{1}{\bar{\alpha}} \sqrt{\frac{\pi}{\bar{\alpha}}}
\ee
and these expressions permit to evaluate the second order derivative (\ref{derivative2}) at the 
condensation point $\mu=\mu_{imp,c}$ ($\nu =0$)  
\be
\label{der2c}
\lim_{\nu\to 0} \frac{\partial^{2}\rho}{\partial\mu ^{2}}=\lim_{\nu\to 0}\left(1+a\frac{\partial\rho_{0}}{\partial\nu}\right)^{-3} 
\frac{\partial^{2}\rho_{0}}{\partial\nu ^{2}} 
\ee
\[ =\frac{\lambda^{6}k_{B}T}{a^3} \lim_{\bar{\alpha}\to 0}\frac{g_{-1/2}(\bar{\alpha})}{[g_{1/2}(\bar{\alpha})]^3} 
 = \frac{\lambda^{6}k_{B}T}{2\pi\, a^3} \quad .\] 

Thus for  $\mu \nearrow \mu_{imp,c} $ one obtains 
\be
\label{vicinity}
\mu - a\rho(T,\mu) \sim  -\frac{\lambda^{6}k_{B}T}{4\pi\, a^2} ( \mu - \mu_{imp,c}  )^2 \quad.
\ee
It follows from the definition of the correlation length $\xi_{imp}$  in equation (\ref{decaymf}) that
\begin{eqnarray}
\label{decaymfg}
\xi_{imp} = \frac{\lambda}{4}\left(-\frac{k_{B}T}{\pi [\mu - a\rho(T,\mu)]} \right)^{1/2} = \nonumber \\  
-\frac{a}{2\lambda^2(\mu_{imp,c}-\mu)} = 
 \frac{\lambda}{2 \zeta (3/2)}\left( 1 - \frac{\mu}{\mu_{imp,c}} \right)^{-1} \quad, 
\end{eqnarray}
where the perfect Bose gas relation $\lambda^3 \rho_{0,c}=\zeta (3/2)$ has been used.

As $\mu$ approaches $\mu_{imp,c}$ from below, the range of the correlation function $\xi_{imp}$ diverges according 
to the power law
\be
\label{divergence mf}
\xi_{imp} =  \frac{\lambda}{2\zeta (3/2)}\left( 1 - \frac{\mu}{\mu_{imp,c}} \right)^{-1},
\ee
with the mean-field critical exponent $\nu_{imp} = 1$.

We now compare the above result with the analogous properties of the decay length $\kappa_{imp}$ of the thermodynamic Casimir force derived by us in \cite{NP2011}. The thermodynamic Casimir force can be presented with the help of the scaling function 
$\Upsilon(z)$ \cite{NP2011,GD2006}
\be
\label{ccf1}
\frac{F(T,D,\mu)}{k_{B}T} = \frac{1}{D^3} \, \left[2 \Upsilon(z)	- z \Upsilon'(z) \right] \quad,
\ee
where $z = D/\kappa_{imp}$ and 
\be 
\label{ccf2}
\Upsilon(z) = - \sum_{n=1}^{\infty} \frac{1 + n z}{\pi n^3} \exp(-n z) \quad,
\ee
such that $\Upsilon(0)=-\zeta(3)/\pi, \Upsilon'(0)=0$, and $\Upsilon(z\gg 1) \approx -1/\pi z\,e^{-z}$. 
However, the expression for the decay length $\kappa_{imp}$ depends on the type of boundary conditions imposed at the walls. For periodic boundary conditions and  near the condensation point the decay length  $\kappa_{imp,per}$ is given by 
\begin{eqnarray}
\label{rangemf}
\kappa_{imp,per}  \, =\, \frac{\lambda \sqrt{\pi}}{\zeta(3/2)}\left( 1 -\frac{\mu}{\mu_{imp,c}} \right)^{-1}
\end{eqnarray} 
We thus find that for an imperfect Bose gas the simple relation $\kappa_{imp,per}=2\sqrt{\pi}\xi_{imp}$ holds confirming the proportionality of these two quantities with their universal $2\sqrt{\pi}$ ratio near the condensation point. These two quantities  diverge according to the same power law with the critical exponent $\nu_{imp}=1$. Similar conclusion holds for the case of Dirichlet ($\kappa_{imp,D}$) and Neumann ($\kappa_{imp,N}$) boundary conditions for which 
$\kappa_{imp,D} = \kappa_{imp,N} = \kappa_{imp,per} /2$.   We note that one could use the correlation length $\xi_{imp}$ instead of the decay length 
$\kappa_{imp}$ as the characteristic length rescaling the distance $D$ in the scaling function $\Upsilon$ in Eq.(\ref{ccf1}). The consequence of this 
new choice for the scaling variable $z'=D/\xi_{imp}$ is that the thermodynamic Casimir force  would be now expressed in terms of the new scaling function 
$\Upsilon_{b.c.}(z')$. Its form depends on the choice of the boundary conditions. \\
The above results show that in the case of periodic boundary conditions the Casimir amplitude $\Delta$ defined as 
$F(T,D,\mu_{imp,c})/(k_{B}T) = 2\,\Delta /D^3$ is equal $\Delta_{imp,per} = -\zeta(3)/\pi$. The same value of the Casimir amplitude has been 
obtained for three-dimensional ideal Bose gas with periodic boundary conditions at the condensation point $\mu_{0,c}=0$, see  
\cite{MZ2006,GD2006}. On the other hand the corresponding value  for the three-dimensional spherical model (sm) with periodic boundary conditions is different: 
$\Delta_{sm,per}=-2\zeta(3)/ (5\pi)$, see \cite{D1998}. In the case of fluctuating Goldstone modes of broken continuous symmetry one has 
$\Delta = -\zeta(3)/(16 \pi)$, see \cite{KG1999}. The extensive discussion of Casimir amplitudes for Robin boundary conditions is presented in \cite{SD2008,DS2011}. 

\section{Concluding comments}

{We have analyzed two strongly fluctuating systems  enclosed by planar walls for which the thermodynamic Casimir forces can be 
explicitly calculated: the perfect and the imperfect Bose gases near their condensation points. 
For each system, the decay length $\kappa$ characterizing the range of the exponentially decaying Casimir force in the one-phase region has been evaluated and compared with the relevant, independently derived  bulk correlation length $\xi$. The decay length $\kappa$ depends on the type of boundary conditions imposed at the walls. 

Upon approaching the condensation point of a perfect Bose gas one finds
\be
\label{prop0}
 \kappa_{0,per} = 2\kappa_{0,D} = 2\kappa_{0,N}
\ee 
 while the  bulk correlation length $\xi_{0}$ which does not depend on boundary conditions turns out to coincide with $\kappa_{0,per}$.
The equality $\xi_{0} = \kappa_{0,per}$ implies that 
both quantities are divergent upon approaching the condensation point according to the same power law
$ \sim (\mu_{0,c}-\mu)^{-\nu_{0}}$, where $\mu_{0,c}=0$, and $\nu_{0}=1/2$. 

Similar proportionality relations involving only boundary conditons dependent numerical coefficients
\be
\label{propimp}
\kappa_{imp,per} = 2\kappa_{imp,D} = 2\kappa_{imp,N} = 2 \sqrt{\pi} \xi_{imp} 
\ee
hold true in the case of the imperfect Bose gas near its condensation point at $\mu_{imp,c} =a \rho_{0,c}$. \\

In order to understand the physical origin of factor $2$ in Eqs(\ref{prop0},\ref{propimp}) one has to go back to the structure of the 
energy spectrum under periodic, Dirichlet, and Neumann boundary conditions. The one-particle energy levels corresponding to the motion 
with momentum perpendicular to the walls under Dirichlet and Neumann  boundary conditions 
for walls separated by distance $D$ are given by
$  \epsilon_{n}^{D,N} = \frac{\hbar^2 n^2}{2m}\left(\frac{\pi}{D}\right)^2$, ($ n=1,2, ... $ for Dirichlet, and $ n=0, 1,2, \cdots $ 
for Neumann). They are equal (for $n \neq 0$) to the energy levels under periodic boundary conditions corresponding to distance $2D$:   $ \epsilon_{n}^{per} = \frac{\hbar^2 n^2}{2m}\left( \frac{2\pi}{2D}\right)^2, \;\;\;n=0,\pm 1,\pm 2, \cdots $. It is rather straightforward to check that this fact implies the following relation for the Casimir excess surface energies: 
$\omega^{<}_{s}(T,2D,\mu)|_{per} = 2 \omega^{<}_{s}(T,D,\mu)|_{D,N}$     
(see Eqs(43),(44), and (54) in ref.[32]). The definition of the Casimir force  
$F(T,D,\mu) = - \frac{\partial \omega^{<}_{s}(T,D,\mu)}{\partial D}$ leads then to simple equality:  
$F_{per}(2D)  =  F_{D,N}(D) $. In the case of exponential decay (see Eqs(\ref{ccf1},\ref{ccf2})) such an equality can hold only if the range of Casimir forces $\kappa$ under periodic boundary conditions is twice as big as in the case of Dirichlet and Neumann boundary conditions. We note that the above relation between the Casimir forces remains valid 
also in the presence of condensate (see Eqs(52, 53) in [32]). Hence, the factor $2$ in Eqs(\ref{prop0}, \ref{propimp}) reflects the simple relation  between the energy spectra. \\
However, in the case of the imperfect Bose gas the value of the corresponding critical exponent differs form that of the perfect Bose gas, and equals 
$\nu_{imp}=1$ (see (\ref{rangemf})) . This fact puts the imperfect Bose gas in a different universality class from the point of view of the behavior of correlations.  In fact, one could expect here some change, as, contrary to $H_{0}$, the Hamiltonian $H_{imp}$ given in (\ref{impham}) is superstable \cite{ZB2001, D1972} implying well defined thermodynamics for any value of the chemical potential. The question of the influence of superstability on thermodynamic Casimir forces has been raised in \cite{MZ2006}, and the present study together with previous work \cite{NP2011} provides a precise answer to it. 
Let us stress once more that the  passage of the bulk density correlation function and the Casimir force from exponential decay to a power law decay at condensation point occurs according to the same mechanism as far as the divergence of the characteristic length  scales is concerned.

\section{Appendix}

We derive here a simple relation between the two-particle density matrices evaluated for the imperfect and perfect Bose gases. 
Our argument is based on the hierarchy equations satisfied by thermodynamic (imaginary time) Green functions \cite{AP2011,FW2003}.

The one-body function $G^{(1)}$ is defined by 
\be
\la{onegreen}
G^{(1)}(\br_1,\tau_1|\br_2,\tau_2) = \langle \rm{T}_\tau [\psi_{\tau_1}(\mathbf{r}_1)\; 
\psi_{\tau_2}^\dagger(\mathbf{r}_2)] \rangle \quad,  
\ee
where $< ... >$ denotes the grand-canonical average   and
\be
\la{anncrea}
\psi_{\tau_{1}}(\br_1) = \exp[\tau_1(H-\mu N)] \psi (\br_1) \exp[-\tau_1(H-\mu N)]  
\ee
\[ \psi_{\tau_2}^\dagger(\mathbf{r}_2)= \exp[\tau_2(H-\mu N)] \psi^\dagger (\br_2 ) \exp[-\tau_2(H-\mu N)] \]
are the imaginary time evolved bosonic annihilation $\psi(\br_1)$ and creation $\psi^\dagger (\br_2 ) $ operators.
In Eq.(\ref{onegreen}) they are ordered by the $\rm{T}_{\tau}$-operator in a chronological order with decreasing times from the left to the right. 
Their commutation relations read
\be
\la{commutation}
[\psi(\mathbf{r}),\psi(\mathbf{r}')] = 0 \; , \;
[\psi^\dagger(\mathbf{r}),\psi^\dagger(\mathbf{r}')] =  0 \; , \; 
[\psi(\mathbf{r}),\psi^\dagger(\mathbf{r}')] = \delta (\mathbf{r}-\mathbf{r}') \quad.
\ee

The system's Hamiltonian has the standard form 
\be
\la{hamiltonian} 
H=-\frac{\hbar^2}{2m} \int \dd \mathbf{r} \; \psi^\dagger(\mathbf{r}) \; \Delta \psi(\mathbf{r}) 
+\frac{1}{2} \int \dd \mathbf{r} \; \dd \mathbf{r}' \; \psi^\dagger(\mathbf{r}')\psi^\dagger(\mathbf{r}) 
\; \Phi(\mathbf{r}-\mathbf{r}') \; \psi(\mathbf{r})\psi(\mathbf{r}') \quad,
\ee
where $\Phi(\mathbf{r}-\mathbf{r}')$ denotes the interparticle potential.  

In order to derive the first hierarchy equation one applies the partial derivative $\partial/\partial \tau_{1}$ to $G^{(1)}(\br_1,\tau_1|\br_2,\tau_2)$. 
The result can be expressed in terms of the two-body Green function
\be
\la{twogreen}
G^{(2)}(\br_1,\tau_1;\br_3,\tau_3|\br_2,\tau_2;\br_4,\tau_4) = \langle \rm{T}_\tau [\psi_{\tau_1}(\mathbf{r}_1)\; \psi_{\tau_3}(\mathbf{r}_3) \;
\psi_{\tau_2}^\dagger(\mathbf{r}_2) \; 
\psi_{\tau_4}^\dagger(\mathbf{r}_4)] \rangle \quad. 
\ee

A straightforward calculation (see ch.7 in \cite{FW2003}, p.172, problem 7.2 ) yields
\begin{multline}
\la{firsthierarchy}
\frac{\partial G^{(1)}}{\partial \tau_1} = \frac{\hbar^2}{2m} 
\Delta_1 G^{(1)} + (\mu - a \rho) \; G^{(1)} + \; \delta (\mathbf{r}_1-\mathbf{r}_2) \; \delta (\tau_1-\tau_2) \\
-\int \dd \mathbf{r}_3 \; \Phi(|\mathbf{r}_3-\mathbf{r}_1|) \left[ G^{(2)}(\br_1, \tau_1,\br_3, \tau_1|\br_2,\tau_2, \br_3,\tau_{1+}) \; \right. \\
\left. - G^{(1)}(\br_1,\tau_1|\br_2,\tau_2))G^{(1)}(\br_3, \tau_1| \br_3,\tau_{1+})\right]\; , 
\end{multline}
where $a=\int \dd \mathbf{r} \; \Phi(r)$, and  $\tau_{1+}=\lim_{0<\epsilon\to 0} (\tau_1+ \epsilon)$.

In the case of an imperfect mean-field gas the potential energy per pair of bosons is a constant
\be
\label{potential}
\Phi(\mathbf{r}-\mathbf{r}')\equiv \frac{a}{V}, 
\ee
inversely proportional to the total volume $V$.
Inserting (\ref{potential}) into (\ref{hamiltonian}) and using the commutation relations (\ref{commutation})
we find  the mean field Hamiltonian 
\be
\la{mfham}
H_{imp} = -\frac{\hbar^2}{2m} \int \dd \mathbf{r} \; \psi^\dagger(\mathbf{r}) \; \Delta \psi(\mathbf{r}) +a \frac{N(N-1)}{2V} \quad,
\ee
where $N$ is the particles number operator 
\be
\la{number}
N=\int \dd \mathbf{r} \; \psi^\dagger(\mathbf{r}) \psi(\mathbf{r}) \quad. 
\ee

For a constant potential (\ref{potential})  the last term on the right hand side of (\ref{firsthierarchy}) takes the form
\begin{eqnarray}
\la{potterm}
-\frac{a}{V}\int \dd \mathbf{r}_3 \;  \left[ G^{(2)}(\br_1, \tau_1;\br_3, \tau_1|\br_2,\tau_2; \br_3,\tau_{1+}) \; \right. 
\nonumber \\
\left.- G^{(1)}(\br_1,\tau_1|\br_2,\tau_2))G^{(1)}(\br_3, \tau_1| \br_3,\tau_{1+})\right]\quad, 
\end{eqnarray}
where the integrand is a short-range integrable function. Thus,  owing to the prefactor $a/V$ 
the whole term vanishes in the thermodynamic limit $V \to \infty$.

In this way we arrive at the conclusion that the one-body Green function $G^{(1)}_{imp}$ of an imperfect Bose gas satisfies 
a {\it closed} equation
\be
\la{firstmf}
\frac{\partial G^{(1)}_{imp}}{\partial \tau_1} = \frac{\hbar^2}{2m} 
\Delta_1 G^{(1)}_{imp} + (\mu - a \rho) \; G^{(1)}_{mf} + \; \delta (\mathbf{r}_1-\mathbf{r}_2) \; \delta (\tau_1-\tau_2) \quad.
\ee
On the other hand, in the case of a perfect Bose gas the corresponding equation reads
\be
\la{firstmf}
\frac{\partial G^{(1)}_{0}}{\partial \tau_1} = \frac{\hbar^2}{2m} 
\Delta_1 G^{(1)}_{0} + \mu  \; G^{(1)}_{0} + \; \delta (\mathbf{r}_1-\mathbf{r}_2) \; \delta (\tau_1-\tau_2) \quad.
\ee
We thus see that the case of an imperfect gas is obtained from that of a perfect gas  by simply replacing the chemical potential
$\mu$ by $\nu = (\mu - a\rho )$ in $G^{(1)}_{0}$. 

Exactly the same situation persists in the case of the two-body Green function (\ref{twogreen}) whose cluster decomposition 
\be
\la{cluster}
G^{(2)}(\br_1,\tau_1;\br_3,\tau_3|\br_2,\tau_2;\br_4,\tau_4) = 
G^{(1)}(\br_1,\tau_1|\br_2,\tau_2) \; G^{(1)}(\br_3,\tau_3|\br_4,\tau_4) 
\ee
\[ + \; G^{(1)}(\br_1,\tau_1|\br_4,\tau_4) \; G^{(1)}(\br_3,\tau_3|\br_2,\tau_2) 
+ G^{(2,T)}(\br_1,\tau_1;\br_3,\tau_3|\br_2,\tau_2;\br_4,\tau_4) \] 
is used to define the truncated function $ G^{(2,T)}$.
A rather complicated structure of the hierarchy equation satisfied by $ G^{(2,T)}$ has been presented in Ref. \cite{AP2011}. 
The important property of this equation is that upon inserting the mean-field potential $a/V$ into Eq.(26) of Ref.\cite{AP2011} 
and taking the thermodynamic limit one arrives again at a closed equation 
\be
\la{secondmf}
\frac{\partial G^{(2,T)}_{imp}}{\partial \tau_2} = -\frac{\hbar^2}{2m} 
\Delta_2 G^{(2,T)}_{imp} - (\mu - a \rho) \; G^{(2,T)}_{imp} \quad.
\ee
The corresponding equation for the perfect Bose gas reads
\be
\la{second0}
\frac{\partial G^{(2,T)}_{0}}{\partial \tau_2} = -\frac{\hbar^2}{2m} 
\Delta_2 G^{(2,T)}_{0} - \mu  \; G^{(2,T)}_{0} \quad.
\ee
We thus conclude that the functions $G^{(2,T)}_{imp}$, and thus also $G^{(2)}_{imp}$, can be obtained by simply replacing $\mu$ by 
$(\mu - a \rho)$ in the corresponding perfect gas functions.  

As the two-particle number density $n_{2}(|\br_1-\br_2|)$ is given by equal time diagonal elements of the two-body Green function
\be
\label{corrgreen}
n_{2}(|\br_{1}-\br_{2}|)
= G^{(2)}(\br_1,\tau_1,\br_2,\tau_1|\br_1,\tau_{1+},\br_2,\tau_{1+}) 
\ee
we also conclude that the correlation function $\chi_{imp}(r)$ of the imperfect Bose gas is equal to the  perfect Bose gas correlation function $\chi_{0}(r)$ evaluated 
at shifted chemical potential $\nu = \mu - a \rho $.
The analysis presented in Section III is based on this observation.

\end{document}